\documentclass[12pt]{article}

\usepackage{graphicx}
\usepackage{amsmath,amssymb}
\usepackage{bm}
\usepackage{graphicx, color}
\usepackage{wrapfig}
\begin{document}
\title{
\begin{flushright}
\ \\*[-80pt] 
\begin{minipage}{0.2\linewidth}
\normalsize
KIAS-Q16001 \\*[50pt]
\end{minipage}
\end{flushright}
{\Large \bf 
Mass Limit for Light Flavon \\
with Residual $Z_3$ Symmetry
\\*[20pt]}}

\author{ 
\centerline{
Yu~Muramatsu$^{1,}$\footnote{E-mail address: yumura@kias.re.kr},
~Takaaki~Nomura$^{1,}$\footnote{E-mail address: nomura@kias.re.kr},~and
~Yusuke~Shimizu$^{2,}$\footnote{E-mail address: yshimizu@kias.re.kr} }
\\*[20pt]
\centerline{
\begin{minipage}{\linewidth}
\begin{center}
$^1${\it \normalsize
School~of~Physics,~KIAS,~Seoul~130-722,~Republic~of~Korea} \\
$^2${\it \normalsize 
Quantum~Universe~Center,~KIAS,~Seoul~130-722,~Republic~of~Korea}
\end{center}
\end{minipage}}
\\*[50pt]}

\date{
\centerline{\small \bf Abstract}
\begin{minipage}{0.9\linewidth}
\medskip 
\medskip 
\small
We present a modified Altarelli and Feruglio $A_4$ model where an additional $A_4$ singlet-prime flavon is introduced. 
In this model, non-zero $\theta _{13}$ is given by this additional $A_4$ singlet-prime flavon which breaks tri-bimaximal mixing. 
In the framework of the supersymmetry with $U(1)_R$ symmetry, 
we obtain vacuum expectation values (VEVs) and VEV alignments of flavons through driving fields. 
It is considered that flavon induces distinctive flavor violating process if flavon mass is light.
Assuming mass of SUSY particles are sufficiently heavy so that the SUSY contributions can be negligible, 
we discuss the flavor violating Yukawa interaction through flavon exchange in the charged lepton sector. 
According to the potential analysis, the VEV of flavon breaks $A_4$ down to $Z_3$ in the charged lepton sector 
and relation among flavon masses is determined. 
Thanks for the residual $Z_3$ symmetry, many lepton flavor violating decay modes are forbidden except for $\tau \rightarrow \mu \mu \bar{e}$ and 
$\tau \rightarrow e e \bar{\mu}$. A mass limit of the flavon from these three-body decay modes is $60$~GeV taking into account 
the current experimental lower bounds at the Belle experiment. In our model, we predict a ratio of the branching ratios 
$\tau \rightarrow \mu \mu \bar{e}$ and $\tau \rightarrow e e \bar{\mu}$ by using known charged lepton masses. 
We also find that the production cross section for the flavon can be $\mathcal{O}(1)$ fb. 
Thus the flavon would be found at the LHC run 2 by searching for 4-tau lepton signal.
\end{minipage}
}

\begin{titlepage}
\maketitle
\thispagestyle{empty}
\end{titlepage}

\section{Introduction}
The neutrino experiments~\cite{An:2012eh}-\cite{Abe:2013hdq} are one of the most attractive experiments for the evidence of the beyond the standard model (SM). 
Actually, the neutrino oscillation experiments provide us that there are two neutrino mass squared differences and two large mixing angles. 
The reactor neutrino experiments also observed non-zero $\theta _{13}$, which is the last mixing angle of the lepton sector~\cite{An:2012eh,Ahn:2012nd,Abe:2014lus}. 
The T2K experiment reported the first stage of the CP violating Dirac phase $\delta _{CP}$ through the electron neutrino appearance 
in a muon neutrino beam~\cite{Abe:2013hdq}. If the neutrinos are Majorana particles, there are Majorana phases which are also sources of the CP violation. 
The neutrino-less double beta ($0\nu \beta \beta $) decay experiments are looking for the evidence of the Majorana particles 
and give the upper-bound of the effective neutrino mass $m_{ee}$. 
Thus the neutrino experiments provide us a new window of the beyond the SM in the theoretical point of view.

The non-Abelian discrete flavor symmetry can easily explain the large mixing of the lepton sector 
e.g. tri-bimaximal mixing (TBM)~\cite{Harrison:2002er,Harrison:2002kp}, which is a simple mixing paradigm in the lepton sector. 
Actually, Altarelli and Feruglio (AF) proposed $A_4$ model of leptons~\cite{Altarelli:2005yp,Altarelli:2005yx} which contains 
new gauge singlet scalar fields, so-called ``flavons" in addition to the $SU(2)$ doublet SM Higgs field. 
There are many authors who study flavor structure to derive TBM by using non-Abelian discrete flavor symmetry not only $A_4$ group 
but also many other groups (See Refs. of~\cite{Ishimori:2010au}-\cite{King:2014nza}.).
After reactor experiments reported non-zero $\theta _{13}$, it is important to study the deviation from TBM precisely~\cite{BhupalDev:2011gi}-\cite{Shimizu:2015tta} 
or study other flavor paradigms, e.g. tri-bimaximal-Cabibbo mixing~\cite{King:2012vj,Shimizu:2012ry}. 

Flavor models using non-Abelian discrete flavor symmetry have not been confirmed in the experimental point of view.
Of course many authors discuss, under the framework of the flavor symmetry models, the lepton flavor violation (LFV) e.g. $\mu \to e\gamma $ 
in addition to the prediction of the Dirac CP phase, Majorana phases, and effective mass of the $0\nu \beta \beta $ decay. 
In Refs.~\cite{Holthausen:2012wz}-\cite{Varzielas:2015sno}, they discussed the mass restriction on the flavons, which are 
related to the SM Higgs, from the LFV and collider physics. However experimental constraints for gauge singlet flavons are not investigated 
although neutrino experimental data can be explained by models with gauge singlet flavons. 
Then, we discuss a mass restriction on the gauge singlet flavons. 
In general, the mass scale of the SM gauge singlet flavons and cutoff scale of the non-Abelian discrete flavor symmetry are taken very high scale. 
Actually, in Refs~\cite{Altarelli:2005yp,Altarelli:2005yx}, if the Dirac neutrino Yukawa couplings are $\mathcal{O}(1)$, 
the cutoff scale should be high scale such as $\mathcal{O}(10^{13})$--$\mathcal{O}(10^{15})$~GeV. 
They assumed that the magnitude of the flavon vacuum expectation values (VEVs) are almost same and tau lepton Yukawa coupling is perturbative. 
Then, the ratio of the flavon VEVs and cutoff scale should be lager than $0.0022$. 
Therefore, the flavon mass is much heavier than the electro weak (EW) scale. 
However this requirement can be relaxed, if we take Dirac neutrino Yukawa couplings to be much less than 
$\mathcal{O}(1)$.\footnote{If we introduce the Froggatt-Nielsen (FN) mechanism~\cite{Froggatt:1978nt} as an additional $U(1)_{FN}$ flavor symmetry, 
the Yukawa couplings can be $\mathcal{O}(1)$.} 
Then the flavon mass can be light without theoretical contradiction. 
Therefore we discuss the lower limit of the flavon mass from the experimental data in our paper.

We present a modified AF $A_4$ model which introduces an additional $A_4$ singlet-prime flavon breaking TBM~\cite{Brahmachari:2008fn}-\cite{Karmakar:2014dva}, 
and calculate a potential of the flavon scalar fields. As well known, we need VEVs of flavons with specific alignments 
in order to obtain the correct masses and mixing angles in the lepton sector. In the framework of the supersymmetry (SUSY) with $U(1)_R$ symmetry, 
we obtain the VEVs of flavons and the alignments of them. Because the SUSY particles have not been found, 
we assume the mass of SUSY particles are sufficiently heavy so that the SUSY contributions can be negligible. 
Then we discuss the flavor violating Yukawa interaction through flavon exchange in the charged lepton sector. 
According to the potential analysis, the VEV of flavon breaks $A_4$ down to $Z_3$ in the charged lepton sector 
and relation among flavon masses is determined. 
Thanks for the residual $Z_3$ symmetry, many lepton flavor violating decay modes are forbidden except for $\tau \rightarrow \mu \mu \bar{e}$ and 
$\tau \rightarrow e e \bar{\mu}$~\cite{Ma:2010gs}. These three-body decay modes are mediated by the flavons.  
Therefore a mass limit of the flavon is $60$~GeV taking into account the current experimental lower bounds at the Belle experiment~\cite{Hayasaka:2010np}. 
In addition, we predict a ratio of the branching ratios $\tau \rightarrow \mu \mu \bar{e}$ and $\tau \rightarrow e e \bar{\mu}$ by using known charged lepton masses. 

In our model, the contribution of the muon $g-2$ is small and constraint from the LEP data~\cite{LEP:2003aa} 
is also less stringent than the constraint given by flavor violating $\tau$ decay. 
Then we discuss the production at the hadron collider through radiation from charged leptons as other candidates for collider signatures. 
We find that the production cross section for the flavon can be $\mathcal{O}(1)$ fb. 
Thus the flavon would be found at the LHC run 2 by searching for 4-tau lepton signal. 

In section~\ref{sec:model}, we show the modified AF $A_4$ model and discuss the potential of the flavons. In section~\ref{sec:LFV}, 
we present the numerical analysis of the flavor physics and collider physics from flavon exchange. 
The section~\ref{sec:summary} is devoted to discussions and summary. 
In appendix~\ref{sec:multiplication-rule}, we show the multiplication rule of the $A_4$ group. 
We show a full scalar potential of the relevant flavon in appendix~\ref{sec:potential-full}.


\section{$A_4$ flavor model}
\label{sec:model}
\begin{table}[b]
\begin{center}
\begin{tabular}{|c|ccccc||c||cccc||ccc|}
\hline
& $l$ & $e_R^c$ & $\mu _R^c$ & $\tau _R^c$ & $\nu _R^c$ & $h_{u,d}$ & $\phi _T $ & $\phi _S$ & $\xi $ & $\xi '$ & $\phi _0^T$ & $\phi _0^S$ & $\xi _0$ \\ 
\hline 
$SU(2)$ & $2$ & $1$ & $1$ & $1$ & $1$ & $2$ & $1$ & $1$ & $1$ & $1$ & $1$ & $1$ & $1$ \\
$A_4$ & $\bf 3$ & $\bf 1$ & $\bf 1''$ & $\bf 1'$ & $\bf 3$ & $\bf 1$ & $\bf 3$ & $\bf 3$ & $\bf 1$ & $\bf 1'$ & $\bf 3$ & $\bf 3$ & $\bf 1$ \\
$Z_3$ & $\omega $ & $\omega ^2$ & $\omega ^2$ & $\omega ^2$ & $\omega ^2$ & $1$ & $1$ & $\omega ^2$ & $\omega ^2$ & $\omega ^2$ & $1$ & $\omega ^2$ & $\omega ^2$ \\
$U(1)_R$ & $1$ & $1$ & $1$ & $1$ & $1$ & $0$ & $0$ & $0$ & $0$ & $0$ & $2$ & $2$ & $2$ \\
\hline 
\end{tabular}
\end{center}
\caption{Assignments of leptons, Higgs, flavons, and driving fields.}
\label{tab:assignment}
\end{table}
In this section, we present a modified AF $A_4$ model 
which introduces an additional $A_4$ singlet-prime flavon~\cite{Shimizu:2011xg}.
Under the $A_4$ group, the left-handed lepton doublet $l=(l_e, l_\mu, l_\tau)$ are assumed to transform
as the triplet, while the right-handed  charged leptons are assigned
to the singlets as $\bf 1$, $\bf 1''$, and $\bf 1'$ for $e_R^c$, $\mu _R^c$, and $\tau _R^c$, respectively.
In Ref.~\cite{Shimizu:2011xg}, the neutrino Majorana masses come from the Weinberg operator~\cite{Weinberg:1979sa}. 
In this paper, we introduce the right-handed neutrinos which are gauge singlet and assigned to the triplet $\nu ^c_R=(\nu _{eR}^c, \nu _{\mu R}^c,\nu ^c_{\tau R})$. 
These right-handed neutrinos can be origin of this Weinberg operator through the seesaw mechanism~\cite{Minkowski:1977sc}-\cite{Schechter:1981cv} in our model. 
$Z_3$ charges are assigned relevantly to the leptons, here we define $\omega ^3=1$. 
On the other hand, the Higgs doublet $h_u$ and $h_d$ are assigned to the $A_4$ trivial singlet. 
We add gauge singlet flavons $\phi _T$, $\phi _S$, $\xi $, and $\xi '$ 
which are assigned to the triplet for $\phi _T=\left (\phi _{T1},\phi _{T2},\phi _{T3}\right )$ and $\phi _S=\left (\phi _{S1},\phi _{S2},\phi _{S3}\right )$ 
and the trivial singlet for $\xi $, and the singlet-prime for $\xi '$ under $A_4$ group, respectively. 
These flavons have different $Z_3$ charges as seen in Table~\ref{tab:assignment}. 
In order to obtain VEVs and VEV alignments, we also add so-called ``driving fields" $\phi _0^T=\left (\phi _{01}^T,\phi _{02}^T,\phi _{03}^T\right )$ 
and $\phi _0^S=\left (\phi _{01}^S,\phi _{02}^S,\phi _{03}^S\right )$ which are assigned to the triplet 
and $\xi _0$ which is assigned to the trivial singlet under $A_4$ group, respectively. 
We can generate the VEV alignments through $F$-terms by coupling flavons to driving fields, 
which carry the $R$ charge $+2$ under $U(1)_R$ symmetry. We also assign $R$ charge $+1$ to the lepton doublets, 
right-handed charged leptons, and right-handed Majorana neutrinos. 
The charge assignments of driving fields are also shown in Table~\ref{tab:assignment}. 
Note that in the original AF model, they introduced $A_4$ trivial singlet flavon $\tilde \xi $ 
which has the same quantum numbers of $\xi $ and was necessary to obtain a non-trivial VEV structure from the minimization of the potential. 
However $A_4$ singlet-prime flavon $\xi '$ gives VEV structure without extra flavon $\tilde \xi $~\cite{Varzielas:2012ai}. 
In these setup, the superpotential for respecting $A_4 \times Z_3$ symmetry 
at the leading order in terms of the $A_4$ cutoff scale $\Lambda $ is written as 
\begin{align}
w&\equiv w_Y +w_d+h.c., \nonumber \\
w_Y&\equiv w_\ell +w_D+w_N, \nonumber \\
w_\ell &=y_e\phi _Tle_R^ch_d/\Lambda +y_\mu \phi _Tl\mu _R^ch_d/\Lambda +y_\tau \phi _Tl\tau _R^ch_d/\Lambda , \nonumber \\
w_D&=y_Dl\nu _R^ch_u, \nonumber \\
w_N&=y_{\phi _S}\phi _S\nu _R^c\nu _R^c+y_\xi \xi \nu _R^c\nu _R^c+y_{\xi '}\xi '\nu _R^c\nu _R^c, \nonumber \\
w_d&\equiv w_d^T+w_d^S, \nonumber \\
w_d^T&=-M\phi _0^T\phi _T+g\phi _0^T\phi _T\phi _T, \nonumber \\
w_d^S&=g_1\phi _0^S\phi _S\phi _S -g_2\phi _0^S\phi _S\xi +g_2'\phi _0^S\phi _S\xi '+g_3\xi _0\phi _S\phi _S-g_4\xi _0\xi \xi , 
\label{eq:superpotential}
\end{align}
where $y$'s are complex Yukawa couplings, $M$ is generally complex mass parameter, 
and $g$'s are trilinear couplings which are also complex 
parameters.\footnote{In order to obtain the positive number of $v_T$, $v_S$, $u$, and $u'$ for Eqs.~(\ref{eq:alignment-vT}) and (\ref{eq:alignment-vS}), 
we take negative sign for several terms in Eq.~(\ref{eq:superpotential}).}
From this superpotential, we discuss the potential analysis in the next subsection.

\subsection{Potential analysis}
\label{sec:potential-analysis}
In this subsection, we discuss the potential for scalar fields including flavons and driving fields. 
Let us write down the superpotential $w_d^T$ and $w_d^S$ in Eq.~(\ref{eq:superpotential}) as 
\begin{align}
w_d^T&=-M\left (\phi _{01}^T\phi _{T1}+\phi _{02}^T\phi _{T3}+\phi _{03}^T\phi _{T2}\right ) \nonumber \\
&+\frac{2g}{3}\left [\phi _{01}^T\left (\phi _{T1}^2-\phi _{T2}\phi _{T3}\right )+\phi _{02}^T\left (\phi _{T2}^2-\phi _{T1}\phi _{T3}\right )
+\phi _{03}^T\left (\phi _{T3}^2-\phi _{T1}\phi _{T2}\right )\right ], \nonumber \\
w_d^S&=\frac{2g_1}{3}\left [\phi _{01}^S\left (\phi _{S1}^2-\phi _{S2}\phi _{S3}\right )+\phi _{02}^S\left (\phi _{S2}^2-\phi _{S1}\phi _{S3}\right )
+\phi _{03}^S\left (\phi _{S3}^2-\phi _{S1}\phi _{S2}\right )\right ] \nonumber \\
&-g_2\left (\phi _{01}^S\phi _{S1}+\phi _{02}^S\phi _{S3}+\phi _{03}^S\phi _{S2}\right )\xi 
+g_2'\left (\phi _{01}^S\phi _{S3}+\phi _{02}^S\phi _{S2}+\phi _{03}^S\phi _{S1}\right )\xi ' \nonumber \\
&+g_3\xi _0\left (\phi _{S1}^2+2\phi _{S2}\phi _{S3}\right )-g_4\xi _0\xi ^2.
\end{align}
Then, the scalar potential is given as
\begin{align}
V&\equiv V_T+V_S, \nonumber \\
V_T&=\sum _i\left |\frac{\partial w_d^T}{\partial \phi _{0i}^T}\right |^2+h.c. \nonumber \\
&=2\left |-M\phi _{T1}+\frac{2g}{3}\left (\phi _{T1}^2-\phi _{T2}\phi _{T3}\right )\right |^2
+2\left |-M\phi _{T3}+\frac{2g}{3}\left (\phi _{T2}^2-\phi _{T1}\phi _{T3}\right )\right |^2 \nonumber \\
&+2\left |-M\phi _{T2}+\frac{2g}{3}\left (\phi _{T3}^2-\phi _{T1}\phi _{T2}\right )\right |^2, \nonumber \\
V_S&=\sum \left |\frac{\partial w_d^S}{\partial X}\right |^2+h.c. \nonumber \\
&=2\left |\frac{2g_1}{3}\left (\phi _{S1}^2-\phi _{S2}\phi _{S3}\right )-g_2\phi _{S1}\xi +g_2'\phi _{S3}\xi '\right |^2
+2\left |\frac{2g_1}{3}\left (\phi _{S2}^2-\phi _{S1}\phi _{S3}\right )-g_2\phi _{S3}\xi +g_2'\phi _{S2}\xi '\right |^2 \nonumber \\
&+2\left |\frac{2g_1}{3}\left (\phi _{S3}^2-\phi _{S1}\phi _{S2}\right )-g_2\phi _{S2}\xi +g_2'\phi _{S1}\xi '\right |^2
+2\left |g_3\left (\phi _{S1}^2+2\phi _{S2}\phi _{S3}\right )-g_4\xi ^2\right |^2,
\label{eq:scalar-potential}
\end{align}
where $X=\phi _{0i}^S,~\xi _0$. 
Therefore, VEV alignment of $\phi _T$ is derived from the condition of the potential minimum ($V_T=0$) in Eq.(\ref{eq:scalar-potential}) as
\begin{equation}
\langle \phi _T\rangle =v_T(1,0,0),\qquad v_T=\frac{3M}{2g},
\label{eq:alignment-vT}
\end{equation}
here $v_T$ is generally complex number because $M$ and $g$ are complex.
Using the VEV and the VEV alignment of Eq.~(\ref{eq:alignment-vT}), we obtain that the charged lepton mass matrix is diagonal.
Then, we can remove the phase of $v_T$ and take $M/g$ as real parameter without loss of generality.
Hereafter we take $M$ and $g$ as real parameters for simplicity. 
On the other hand, VEV alignment of $\phi _S$ and VEVs of $\xi $ and $\xi '$ 
are derived from the condition of the potential minimum ($V_S=0$) in Eq.(\ref{eq:scalar-potential}) as
\begin{equation}
\langle \phi _S\rangle =v_S(1,1,1),\quad \langle \xi \rangle =u,\quad \langle \xi '\rangle =u',
\quad v_S^2=\frac{g_4}{3g_3}u^2,\quad u'=\frac{g_2}{g_2'}u.
\label{eq:alignment-vS}
\end{equation}
Therefore, we can take the VEVs $u$ and $u'$ as arbitrary numbers.
Using the VEVs and the VEV alignment of Eq.~(\ref{eq:alignment-vS}), the neutrino mass matrix derives the lepton mixing 
by taking an additional rotation of $1$-$3$ generations of neutrinos in the TBM. 
Then, we obtain the non-zero $\theta _{13}$ which comes from $A_4$ singlet flavon VEV ratio $u'/u$ (See Ref.~\cite{Brahmachari:2008fn}-\cite{Karmakar:2014dva}.).

Before closing this subsection, we discuss the $A_4$ breaking patterns.
$A_4$ is the symmetry group of a tetrahedron or even permutation of four elements. 
The number of elements is $12$. The irreducible representations of $A_4$ are $\bf 1$, $\bf 1'$, $\bf 1''$, and $\bf 3$. 
Also $A_4$ can be defined as the group generated by two elements $S$ and $T$ which satisfy the algebraic relations as
\begin{equation}
S^2=T^3=\left (ST\right )^3={\bf 1}.
\end{equation}
On the one-dimensional representations, these generators are represented by
\begin{equation}
\begin{array}{rcl}
\bf 1: & S=1, & T=1, \\
\bf 1': & S=1, & T=e^{4\pi i/3}\equiv \omega ^2, \\
\bf 1'': & S=1, & T=e^{2\pi i/3}\equiv \omega .
\end{array}
\label{eq:one-dimensional-rep}
\end{equation}
On the three-dimensional representation, these generators are represented by  
\begin{equation}
{\bf 3}:\quad T=
\begin{pmatrix}
1 & 0 & 0 \\
0 & \omega ^2 & 0 \\
0 & 0 & \omega 
\end{pmatrix},\quad S=\frac{1}{3}
\begin{pmatrix}
-1 & 2 & 2 \\
2 & -1 & 2 \\
2 & 2 & -1
\end{pmatrix}.
\label{eq:three-dimensional-rep}
\end{equation}
After $A_4$ is broken by taking the VEVs and the VEV alignments of flavons $\phi _T$ and $\phi _S$, 
there are two breaking patterns $A_4\to G_T$ and $A_4\to G_S$ where $G_T$ and $G_S$ are subgroups of $A_4=\left (Z_2\times Z_2\right )\rtimes Z_3$. 
$\langle \phi _T\rangle =v_T\left (1,0,0\right )$ breaks $A_4$ down to $G_T=Z_3$, while 
$\langle \phi _S\rangle =v_S\left (1,1,1\right )$ breaks $A_4$ down to $G_S=Z_2$.
Therefore, the LFV is restricted by residual $Z_3$ symmetry in the charged lepton sector. 
Hereafter, we focus on flavon $\phi _T$ which couples to the charged lepton sector, 
since we will discuss LFV and collider physics in section~\ref{sec:LFV}. 
In the next subsection, we discuss the mass of flavon $\phi _T$.

\subsection{Mass of the flavon}
In this subsection, we discuss the mass of flavon $\phi _T$ which couples to the charged lepton sector.
We expand the flavon field around the VEV $v_T$ as 
\begin{equation}
\phi _T=\left (\phi _{T1},\phi _{T2},\phi _{T3}\right )\rightarrow \left (v_T+\varphi _{T1},\varphi _{T2},\varphi _{T3}\right ),
\label{eq:expanding-around-VEV}
\end{equation}
where $\varphi _{Ti}$ are complex scalar fields. 
Then, we rewrite the scalar potential $V_T$ in Eq.~(\ref{eq:scalar-potential}) as 
$V_T=V_T^{\text{mass}}+(\text{other terms})$,\footnote{In appendix~\ref{sec:potential-full}, we show the full scalar potential $V_T$.} 
and $V_T^{\text{mass}}$ is the mass term of flavon $\phi _T$ as
\begin{equation}
V_T^{\text{mass}}=2M^2\left (\left |\varphi _{T1}\right |^2+4\left |\varphi _{T2}\right |^2+4\left |\varphi _{T3}\right |^2\right ),
\end{equation} 
here we eliminate $g$ by using Eq.~(\ref{eq:alignment-vT}). 
Therefore, masses of scalar fields $m_{\varphi _{Ti}}$ 
are obtained as 
\begin{equation}
(m_{\varphi _{T1}}^2,m_{\varphi _{T2}}^2,m_{\varphi _{T3}}^2)=(2M^2,8M^2,8M^2),
\end{equation}
and the scalar fields $\varphi _{Ti}$ do not mix each other in the mass term.
Taking into account these masses, we discuss the flavor phenomenology and collider physics in the next section.


\section{Flavor phenomenology and collider physics \\ from flavon exchange}
\label{sec:LFV}
In this section we discuss flavor phenomenology and collider physics from flavon exchange.
First of all, we assumed masses of SUSY particles to be heavy because we have not found any SUSY particles 
so that the SUSY contributions can be negligible.
Then we show flavon Yukawa interactions in the charged lepton sector. 
Next we show flavor physics from flavon exchange and constraint for flavon mass.
Finally, we show some predictions for flavor phenomenology and collider physics.

\subsection{Flavon Yukawa interactions in charged lepton sector}
In our model SM Yukawa interactions and flavon Yukawa interactions 
in the charged lepton sector come from following Lagrangian in Eq.~(\ref{eq:superpotential}); 
\begin{equation}
\mathcal{L}_\ell =
y_e\left (\phi _T\bar l\right )e_Rh_d/\Lambda +
y_\mu \left (\phi _T\bar l\right )'\mu _R h_d/\Lambda +
y_\tau \left (\phi _T\bar l\right )''\tau _Rh_d/\Lambda +h.c.,
\end{equation}
where we use the same notation for superfields and SM fields. 
After expanding flavon field $\phi _T$ around VEV $v_T$ in Eq.~(\ref{eq:expanding-around-VEV}) and 
taking the VEV of $SU(2)$ doublet Higgs $h_d$ as $v_d$, 
Lagrangian for charged lepton mass terms $\mathcal{L}_{\ell }^{\text{mass}}$ is written as
\begin{align}
\mathcal{L}_{\ell }^{\text{mass}}&=
\begin{pmatrix}
\bar e_L & \bar \mu _L& \bar \tau _L
\end{pmatrix}
\begin{pmatrix}
\frac{y_e v_d}{\Lambda }v_T & 0 & 0 \\
0 & \frac{y_\mu v_d}{\Lambda }v_T & 0 \\
0 & 0 & \frac{y_\tau v_d}{\Lambda }v_T
\end{pmatrix}
\begin{pmatrix}
e_R \\
\mu _R \\
\tau _R
\end{pmatrix}+h.c. \nonumber \\
&\equiv 
\begin{pmatrix}
\bar e_L & \bar \mu _L& \bar \tau _L
\end{pmatrix}
\begin{pmatrix}
m_e & 0 & 0 \\
0 & m_\mu & 0 \\
0 & 0 & m_\tau 
\end{pmatrix}
\begin{pmatrix}
e_R \\
\mu _R \\
\tau _R
\end{pmatrix}+h.c..
\end{align}
In our model, charged leptons in the interaction basis is equal to those in the mass basis. 
Therefore, there is no mixing in the charged lepton sector in the leading level.
Let us discuss the Lagrangian of the charged lepton and flavon interaction which induces flavon Yukawa interactions in the charged lepton sector 
$\mathcal{L}_{\ell }^{\text{FY}}$ as 
\begin{align}
\mathcal{L}_{\ell }^{\text{FY}}&=
\begin{pmatrix}
\bar e_L & \bar \mu _L& \bar \tau _L
\end{pmatrix}
\begin{pmatrix}
\frac{m_e}{v_T} & 0 & 0 \\
0 & \frac{m_\mu}{v_T} & 0 \\
0 & 0 & \frac{m_\tau}{v_T}
\end{pmatrix}
\begin{pmatrix}
e_R \\
\mu _R \\
\tau _R
\end{pmatrix}
\varphi _{T1} \nonumber \\
&+
\begin{pmatrix}
\bar e_L & \bar \mu _L& \bar \tau _L
\end{pmatrix}
\begin{pmatrix}
0 & \frac{m_\mu }{v_T} & 0 \\
0 & 0 & \frac{m_\tau }{v_T} \\
\frac{m_e}{v_T} & 0 & 0
\end{pmatrix}
\begin{pmatrix}
e_R \\
\mu _R \\
\tau _R
\end{pmatrix}
\varphi _{T2} \nonumber \\
&+
\begin{pmatrix}
\bar e_L & \bar \mu _L& \bar \tau _L
\end{pmatrix}
\begin{pmatrix}
0 & 0 & \frac{m_\tau }{v_T} \\
\frac{m_e}{v_T} & 0 & 0 \\
0 & \frac{m_\mu }{v_T} & 0
\end{pmatrix}
\begin{pmatrix}
e_R \\
\mu _R \\
\tau _R
\end{pmatrix}
\varphi _{T3}+h.c..
\end{align}
We find that $\varphi_{T1}$ 
exchange does not induce flavor violation, 
while the other flavon exchanges induce flavor violation. 
The most interesting feature of these flavon interactions is that 
couplings are almost fixed by charged lepton masses 
except for $A_4$ triplet flavon VEV $v_T$. 

Next, we discuss the lepton radiative flavor violating decays $\mu \rightarrow e \gamma$, $\tau \rightarrow \mu \gamma$, and $\tau \rightarrow e \gamma$. 
In the charged lepton sector, because we take the VEV alignment of $\phi _T$ as $v_T \left (1,0,0\right )$, $A_4$ breaks down to $G_T=Z_3$,
which we discussed in section~\ref{sec:potential-analysis}. 
Then flavons $\varphi _{Ti}$ are transformed by Eq.~(\ref{eq:three-dimensional-rep}) as
\begin{equation}
\varphi _{T1}\rightarrow \varphi _{T1},\quad \varphi _{T2}\rightarrow \omega ^2\varphi _{T2},\quad \varphi _{T3}\rightarrow \omega \varphi _{T3},
\end{equation}
under the residual symmetry $G_T=Z_3$. 
The left-handed charged leptons are transformed by Eq.~(\ref{eq:three-dimensional-rep}) as
\begin{equation}
e_L\rightarrow e_L,\quad \mu _L\rightarrow \omega \mu _L,\quad \tau _L\rightarrow \omega ^2\tau _L.
\end{equation}
On the other hand, the right-handed charged leptons are transformed by Eq.~(\ref{eq:one-dimensional-rep}) as
\begin{equation}
e_R\rightarrow e_R,\quad \mu _R\rightarrow \omega \mu _R,\quad \tau _R\rightarrow \omega ^2\tau _R.
\end{equation}
Therefore many lepton flavor violating decay modes are 
forbidden by the residual symmetry $G_T=Z_3$ e.g. $\mu \to e\gamma $ except for several lepton flavor violating three-body decays. 
The muon $g-2$ obtains contribution from one-loop diagram where only $\varphi_{T1}$ propagates inside loop without flavor change.
However the contribution is small since corresponding Yukawa coupling constant is $m_\mu/ v_T$. 
We thus discuss the lepton flavor violating three-body decay in the next subsection. 


\subsection{Lepton flavor violating three-body decay}

In our model many lepton flavor violating three-body decay modes are 
forbidden by the residual symmetry $G_T=Z_3$.
Dominant flavor changing decay modes are $\tau \rightarrow \mu \mu \bar{e}$ and 
$\tau \rightarrow e e \bar{\mu}$~\cite{Ma:2010gs}.
These decay modes are induced by flavon exchange at tree-level and these 
branching ratios are given by
\begin{align}
\text{BR}\left (\tau \rightarrow \mu \mu \bar{e} \right ) &=
\tau _{\tau }\frac{m_{\tau }^5}{3072 \pi ^3}\left (
\left | \frac{m_{\tau }m_{\mu }}{v_T^2 m_{\varphi _{T2}}^2} \right |^2 + 
\left | \frac{m_{\mu }m_{e}}{v_T^2 m_{\varphi _{T3}}^2} \right |^2
\right ) \nonumber \\
&\simeq \frac{2.9 \times 10^6~\text{GeV}^8}{v_{T}^4 (2 \sqrt{2}M)^4},\nonumber \\
\text{BR}\left (\tau \rightarrow e e \bar{\mu } \right ) &=
\tau _{\tau }\frac{m_{\tau }^5}{3072 \pi ^3}\left (
\left | \frac{m_{\mu }m_{e}}{v_T^2 m_{\varphi _{T2}}^2} \right |^2 + 
\left | \frac{m_{\tau }m_{e}}{v_T^2 m_{\varphi _{T3}}^2} \right |^2
\right ) \nonumber \\
&\simeq \frac{68~\text{GeV}^8}{v_{T}^4 (2 \sqrt{2}M)^4},
\end{align}
where $\tau_{\tau}$ is the lifetime of tau lepton and we use the magnitude of the charged lepton masses by the particle data group (PDG)~\cite{Agashe:2014kda}. 
In our model the decay width for $\tau \rightarrow \mu \mu \bar{e}$ is 
much larger than that for $\tau \rightarrow e e \bar{\mu}$.
Current experimental lower bounds at the Belle experiment are  
$\text{BR}\left( \tau \rightarrow \mu \mu \bar{e} \right) < 1.7 
\times 10^{-8}$ and 
$\text{BR}\left( \tau \rightarrow e e \bar{\mu} \right) < 1.5 
\times 10^{-8}$~\cite{Hayasaka:2010np}.
To realize these constraints, $2 \sqrt{2}M \gtrsim 60~\text{GeV}$ is required 
when we assume that VEV $v_T$ is equal to flavon masses $m_{\varphi _{T2}}$, $m_{\varphi _{T3}}$.
In addition, we can predict a ratio of these branching ratios $r$ by using known 
charged lepton masses as 
\begin{equation}
r=\frac{\text{BR}\left (\tau \rightarrow \mu \mu \bar{e} \right )}{\text{BR}\left (\tau \rightarrow e e \bar{\mu} \right )}
=\frac{m_{\mu }^2\left (m_{\tau }^2+m_{e}^2\right )}{m_{e}^2\left (m_{\mu }^2+m_{\tau }^2\right )}\simeq \frac{m_\mu ^2}{m_e^2},
\end{equation}
because of $m_{\varphi_{T2}}=m_{\varphi_{T3}}$. Then the ratio of $\text{BR}\left (\tau \rightarrow \mu \mu \bar{e} \right )$ and 
$\text{BR}\left (\tau \rightarrow e e \bar{\mu }\right )$ is $r\simeq 4.3\times 10^4$.
As a result of the calculations, we find that flavon mass can be very light. Therefore, we discuss the flavon collider phenomenology in the next subsection.

\subsection{Flavon collider phenomenology}
\begin{table}[h]
\begin{center}
\begin{tabular}{cccccc} \hline \hline
Final state & $\varphi _{T1} \tau \bar \tau $ & $\varphi _{T2} \tau \bar \mu$ & $\varphi _{T3} \tau \bar e$ & $\varphi _{T2} \tau \bar \nu_ \mu$ & $\varphi _{T3} \tau \bar \nu_e$ \\
Cross section [fb] & 0.59 & 0.017 & 0.017 & 0.040 & 0.040 \\ \hline \hline 
\end{tabular}
\caption{ Dominant flavon production cross sections at the LHC 14~TeV where $v_T = 2m_{\varphi_{T1}} = m_{\varphi_{T2}} = m_{\varphi_{T3}} = 65$~GeV is adopted. 
The values of cross sections are sum of shown final states and its charge conjugation.}
\label{tab:CX}
\end{center}
\end{table}
After obtaining the constraint on the flavon mass from flavor violating lepton decay, we discuss collider physics of the flavon
where the flavon exchanging and production processes could be experimentally tested since the flavon can be light as $m_{\varphi_{Ti}} \simeq \mathcal{O}(100)$~GeV.

We first discuss constraints on the flavon mass from the t-channel processes $e \bar e \to \mu \bar \mu$ and $\tau \bar \tau$ at the LEP experiment.
The relevant 4-Fermi interactions are obtained via $\varphi_{T2}$, $\varphi_{T3}$ exchange such that 
\begin{equation}
\label{eq:eff}
\mathcal{L}_{\rm 4-fermi} \supset \frac{m_\mu ^2 }{ 64 M^4} (\bar \mu _R e_L)(\bar e_L \mu _R) + \frac{m_\tau ^2 }{ 64 M^4} (\bar \tau _R e_L)(\bar e_L \tau _R) + h.c., 
\end{equation}
where $m_{\varphi_{T2}} = m_{\varphi_{T3}} = 2 \sqrt{2} M$ and $v_T = 2 \sqrt{2} M$ is adopted, and 
we only show the interactions which induce the processes $e \bar e \to \mu \bar \mu$ and $e \bar e \to \tau \bar \tau$. 
We thus find that $e \bar e \to \tau \bar \tau$ process is dominant.
Then we obtain the constraint on the flavon mass from the LEP data~\cite{LEP:2003aa};
\begin{equation}
(2 \sqrt{2} M)^2 \gtrsim 620 m_\tau ~{\rm GeV}. 
\end{equation}
Therefore this constraint requires $2 \sqrt{2} M \gtrsim 33$~GeV which is less stringent than the constraint given by flavor violating $\tau$ decay in previous subsection.

Flavons $\varphi_{Ti}$ can be produced at the hadron collider through radiation from charged leptons, 
i.e. $p p \to \varphi_{Ti} \bar{\ell} \ell'$ and $p p \to \varphi_{Ti} \bar{\ell} \nu (\ell \bar \nu)$ where $\ell = e, \mu$ and $\tau$.
Taking into account the flavon-lepton Yukawa coupling proportional to $m_\tau$, 
the dominant final states in flavon production processes are summarized in the first low of Table.~\ref{tab:CX}.
Produced flavons then decay into lepton pair where the dominant decay modes are $\varphi_{T1} \to \tau \bar \tau$, $\varphi_{T2} \to \mu \bar \tau$ and $\varphi_{T3} \to e \bar \tau $.
Thus the signals are four-leptons including at least two $\tau$ leptons. 
The production cross sections are calculated using CalcHEP~\cite{Belyaev:2012qa} with {\tt CTEQ6L} PDF~\cite{Nadolsky:2008zw}, 
which are shown in the second low of the Table.~\ref{tab:CX}.
Here we adopted $v_T = 2m_{\varphi _{T1}}=m_{\varphi_{T2}}=m_{\varphi_{T3}} = 65$~GeV which is close to the lower limit from flavor violating lepton decay.
We find that the $\varphi_{T1}$ production cross section can be $\mathcal{O}(1)$ fb while those of $\varphi_{T2}$, $\varphi_{T3}$ are $\mathcal{O}(10^{-2})$ fb 
since the mass of $\varphi_{T1}$ is half of the others. 
Thus $\varphi_{T1}$ would be found at the LHC run 2 by searching for 4-tau lepton signal since SM background is not large. 
Moreover $\varphi_{T2}$, $\varphi_{T3}$ provide peak of invariant mass distribution for $\mu \tau$ and $e \tau$ pair respectively 
which are significant signal of flavor violating interaction and SM background also could be highly reduced with relevant kinematical cuts.
Thus $\varphi_{T2}$, $\varphi_{T3}$ will be also important target at the High-Luminosity LHC, 
which could be tested with large amount of integrated luminosity although the production cross sections are small.
Furthermore the flavon-lepton Yukawa interactions can be tested at the lepton colliders like ILC~\cite{Behnke:2013xla}-\cite{Behnke:2013lya} 
where detailed analysis is left as future work.

We can calculate these cross sections as a function of the product of the flavon mass and the VEV because other couplings are determined by charged lepton masses. 
Branching ratio of the lepton flavor violating three-body decay modes is also a function of same one. 
Then, we can predict relation between collider signature and flavor physics. 
Therefore if we measure one of them, the rest one can be signature of our model.

\section{Discussions and Summary}
\label{sec:summary}
Flavor models which introduce gauge singlet flavons using non-Abelian discrete flavor symmetry have not been confirmed in the experimental point of view. 
In general, the mass scale of the SM gauge singlet flavons and cutoff scale of the non-Abelian discrete flavor symmetry are assumed to be very high scale because 
many authors take Yukawa couplings to be $\mathcal{O}(1)$ so that the flavon masses are much heavier than the EW scale. 
However this requirement can be relaxed, if we take Dirac neutrino Yukawa couplings to be much less than $\mathcal{O}(1)$ in our model. 
Then the flavon mass can be light without theoretical contradiction. 
Therefore we discussed the lower limit of the flavon mass from the experimental data in our paper.

We presented the modified AF $A_4$ model which introduces an additional $A_4$ singlet-prime flavon breaking TBM, 
and calculated the potential of the flavon scalar fields. 
Note that sometimes the contributions from the next-to-leading order are added to realize non-zero $\theta _{13}$. 
However we ignored the contributions from the next-to-leading order because the non-zero $\theta _{13}$ has been already derived by the $A_4$ singlet flavon VEV ratio $u'/u$ 
and such contributions can be eliminated in specific UV completions of the flavor models in Refs.~\cite{Varzielas:2012ai,Varzielas:2010mp}. 
As well known, we need the VEVs of flavons with specific alignments 
in order to obtain the correct masses and mixing angles in the lepton sector. In the framework of the SUSY with $U(1)_R$ symmetry, 
we obtained the VEVs of flavons and the alignments of them. Because the SUSY particles have not been found, 
we assume the mass of SUSY particles are sufficiently heavy so that the SUSY contributions can be negligible. 
Then we discussed the flavor violating Yukawa interaction through flavon exchange in the charged lepton sector. 
According to the potential analysis, the VEV of flavon $\phi _T$ breaks $A_4$ down to $G_T=Z_3$ in the charged lepton sector 
and the masses of flavons $m_{\varphi _{T2}}$ and $m_{\varphi _{T3}}$ are same and twice as heavy as $m_{\varphi _{T1}}$. 
Thanks for the residual $Z_3$ symmetry, many lepton flavor violating decay modes are forbidden except for $\tau \rightarrow \mu \mu \bar{e}$ and 
$\tau \rightarrow e e \bar{\mu}$. These three-body decay modes are mediated by flavons $\varphi _{T2}$ or $\varphi _{T3}$. 
Therefore the mass limit of flavons $\varphi _{T2}$, $\varphi _{T3}$ is $60$~GeV taking into account the current experimental lower bounds at the Belle experiment. 
Then if we assume that the magnitude of the flavon VEV is same as the mass of flavon such as $v_T=2\sqrt{2}M$, tau lepton Yukawa coupling $y_\tau $ is $\mathcal{O}(1)$, 
and $\tan \beta =3$, the cutoff scale $\Lambda $ should be at least $\mathcal{O}(10)$~TeV to realize tau lepton mass. 
Therefore we should take $\tan \beta $ to be small and we will find the new physics in the near future. 
In addition, we predicted the ratio of the branching ratios $\tau \rightarrow \mu \mu \bar{e}$ and $\tau \rightarrow e e \bar{\mu}$ such as $r\simeq 4.3\times 10^4$
by using known charged lepton masses. 

The contribution of the muon $g-2$ is small and constraint from the LEP data 
is also less stringent than the constraint given by flavor violating $\tau$ decay in our model. 
Then we discussed the production at the hadron collider through radiation from charged leptons 
as other candidates for collider signatures. 
We found that the $\varphi_{T1}$ production cross section can be $\mathcal{O}(1)$ fb while those of $\varphi_{T2}$, $\varphi_{T3}$ are $\mathcal{O}(10^{-2})$ fb 
since the mass of $\varphi_{T1}$ is half of the others. 
Thus $\varphi_{T1}$ would be found at the LHC run 2 by searching for 4-tau lepton signal.
The $\varphi_{T2}$, $\varphi_{T3}$ provide peak of invariant mass distribution for $\mu \tau$ and $e \tau$ pair respectively 
which are significant signal of flavor violating interaction and SM background also could be highly reduced with relevant kinematical cuts. 
Thus $\varphi_{T2}$, $\varphi_{T3}$ will be also important target at the High-Luminosity LHC, 
which could be tested with large amount of integrated luminosity although the production cross sections are small. 
Furthermore the flavon-lepton Yukawa interactions can be tested at the lepton colliders like ILC. 

In our model, we can predict relation between flavor physics and collider signature because flavor symmetry fixes many couplings and 
the residual $Z_3$ symmetry makes flavon mass limit light as accessible in collider search. 
Therefore if we measure one of them, the rest one can be signature of our model. 
In other flavor models, flavon mass limit is not light as accessible in collider search if there is no residual symmetry. 
On the other hand, there are rich signatures of flavor physics, then we can predict many relations between these signatures. 
Therefore it is important to study phenomenology from flavon exchange in flavor models. 

\vspace{0.5cm}
\noindent
{\bf Acknowledgement}

YM is supported in part by National Research Foundation of Korea (NRF) Research Grant NRF- 2015R1A2A1A05001869.


\appendix 

\section{Multiplication rule of $A_4$ group}
\label{sec:multiplication-rule}
In this appendix A, we show the multiplication of $A_4$ group.
The multiplication rule of the triplet is 
written as follow;
\begin{align}
\begin{pmatrix}
a_1\\
a_2\\
a_3
\end{pmatrix}_{\bf 3}
\otimes 
\begin{pmatrix}
b_1\\
b_2\\
b_3
\end{pmatrix}_{\bf 3}
&=\left (a_1b_1+a_2b_3+a_3b_2\right )_{\bf 1} 
\oplus \left (a_3b_3+a_1b_2+a_2b_1\right )_{{\bf 1}'} \nonumber \\
& \oplus \left (a_2b_2+a_1b_3+a_3b_1\right )_{{\bf 1}''} \nonumber \\
&\oplus \frac13
\begin{pmatrix}
2a_1b_1-a_2b_3-a_3b_2 \\
2a_3b_3-a_1b_2-a_2b_1 \\
2a_2b_2-a_1b_3-a_3b_1
\end{pmatrix}_{{\bf 3}}
\oplus \frac12
\begin{pmatrix}
a_2b_3-a_3b_2 \\
a_1b_2-a_2b_1 \\
a_1b_3-a_3b_1
\end{pmatrix}_{{\bf 3}\ .}
\end{align}
More details are shown in the review~\cite{Ishimori:2010au}-\cite{King:2014nza}.

\section{Full scalar potential $V_T$}
\label{sec:potential-full}
We show the full potential $V_T$ as follow;
\begin{align}
V_T&=2M^2\left (\left |\varphi _{T1}\right |^2+4\left |\varphi _{T2}\right |^2+4\left |\varphi _{T3}\right |^2\right ) \nonumber \\
&+\frac{2M^2}{v_T}\Big [\left (\varphi _{T1}+\varphi _{T1}^*\right )\left (\left |\varphi _{T1}\right |^2+2\left |\varphi _{T2}\right |^2+2\left |\varphi _{T3}\right |^2\right )
-\left (\varphi _{T1}\varphi _{T2}^*\varphi _{T3}^*+\varphi _{T1}^*\varphi _{T2}\varphi _{T3}\right ) \nonumber \\
&\hspace{1.5cm}-2\left (\varphi _{T2}^{~~2}\varphi _{T3}^*+\varphi _{T2}^{*~2}\varphi _{T3}\right )
-2\left (\varphi _{T2}\varphi _{T3}^{*~2}+\varphi _{T2}^*\varphi _{T3}^{~~2}\right )\Big ] \nonumber \\
&+\frac{2M^2}{v_T^2}\Big [\left |\varphi _{T1}\right |^4+\left |\varphi _{T2}\right |^4+\left |\varphi _{T3}\right |^4
+\left |\varphi _{T1}\right |^2\left |\varphi _{T2}\right |^2+\left |\varphi _{T2}\right |^2\left |\varphi _{T3}\right |^2
+\left |\varphi _{T3}\right |^2\left |\varphi _{T1}\right |^2 \nonumber \\
&\hspace{1.5cm}-\left (\varphi _{T1}^{~~2}\varphi _{T2}^*\varphi _{T3}^*+\varphi _{T1}^{*~2}\varphi _{T2}\varphi _{T3}\right ) \nonumber \\
&\hspace{1.5cm}-\left (\varphi _{T1}\varphi _{T2}^{*~2}\varphi _{T3}+\varphi _{T1}^*\varphi _{T2}^{~~2}\varphi _{T3}^*\right )
-\left (\varphi _{T1}\varphi _{T2}\varphi _{T3}^{*~2}+\varphi _{T1}^*\varphi _{T2}^*\varphi _{T3}^{~~2}\right )\Big ],
\end{align}
where we eliminate $g$ by using Eq.~(\ref{eq:alignment-vT}).

\newpage


\end{document}